\documentclass[aps,superscriptaddress,prl,twocolumn]{revtex4-1}
\usepackage{amsmath}
\usepackage{graphicx}
\usepackage{color}
\usepackage{amssymb}
\usepackage[hidelinks]{hyperref}

\begin{document}

\title{Realization of magneto-optical near-zero index medium by an
  unpaired Dirac point}

\author{Xin Zhou}
\affiliation{Division of Physics and Applied Physics, School of Physical and Mathematical Sciences, \\
Nanyang Technological University, Singapore 637371, Singapore}

\author{Daniel Leykam}
\affiliation{Center for Theoretical Physics of Complex Systems, Institute for Basic Science (IBS), Daejeon 34126, Republic of Korea}

\author{A.~B.~Khanikaev}
\affiliation{Department of Physics, Queens College and Graduate Center of The City University of New York, \\ Queens, New York 11367, USA}

\author{Y.~D.~Chong}
\affiliation{Division of Physics and Applied Physics, School of Physical and Mathematical Sciences, \\
Nanyang Technological University, Singapore 637371, Singapore}
\affiliation{Centre for Disruptive Photonic Technologies, Nanyang Technological University, Singapore 637371, Singapore}

\email{yidong@ntu.edu.sg}

\date{\today}

\begin{abstract}
We realize an unpaired Dirac cone at the center of the first Brillouin
zone, using a gyromagnetic photonic crystal with broken square
sub-lattice symmetry and broken time reversal symmetry.  The behavior
of the Dirac modes can be described by a gyromagnetic effective medium
model with near-zero refractive index, and Voigt parameter near unity.
When two domains are subjected to opposite magnetic biases, there
exist unidirectional edge states along the domain wall.  This
establishes a link between topological edge states and the surface
waves of homogenous magneto-optical media.
\end{abstract}

\maketitle

In two-dimensional (2D) lattices, Dirac points are linear
band-crossing points associated with emergent relativistic
two-component particles.  Since the work of Haldane
\cite{haldane1988}, they have been recognized as the elementary
``building blocks'' for topological phases, occurring at the
boundaries separating parametric domains of different band topology
\cite{hasan2010, Qi2011}.  If time-reversal symmetry ($\mathcal{T}$)
is unbroken, Dirac points come in pairs~\cite{Nielsen1981a,
  Nielsen1981b}; in triangular and honeycomb lattices such as
graphene, the pairs are pinned to high-symmetry points at the corners
of the Brillouin zone~\cite{haldane1988,Semenoff1984}.  By breaking
$\mathcal{T}$ and lattice symmetries, however, these constraints can
be relaxed.

This Letter describes an experimentally realizable
$\mathcal{T}$-broken photonic crystal of gyromagnetic cylinders, whose
bandstructure exhibits an unpaired Dirac point at the center of the
Brillouin zone ($\Gamma$).  The unpaired Dirac point causes the
photonic crystal to act as an effective electromagnetic medium with
two interesting properties: near-zero (NZ) refractive index
\cite{EnghetaReview2017} and strongly enhanced magneto-optical
activity.  Using this effective medium description, we are able to
establish a link between topological edge states and surface waves in
magneto-optical NZ media, two phenomena that had previously been
conceptually distinct.  Unlike the more familiar case of paired Dirac
points in triangular or honeycomb photonic crystals, the Dirac modes
in our system can be said to act as an effective medium because they
interact isotropically with current sources, and there does not exist
another valley for them to scatter into.

A number of studies have established the remarkable fact that photonic
crystals with conical band-crossing points can act as NZ index
effective media \cite{CT2011, CT2012, dong2015, Ashraf2016,
  Jahani2016}.  Intuitively, this is due to the vanishing of the phase
velocity near a band-crossing point at $\Gamma$.  This provides a
practical method for using all-dielectric photonic structures to
realize NZ index media, which have numerous exotic capabilities such
as geometry-insensitive wave-guiding \cite{EnghetaReview2017}.
Although the type of band-crossing point utilized in these schemes
\cite{CT2011, CT2012, dong2015, Ashraf2016, Jahani2016,xiao2017} is
sometimes called ``Dirac-like'', due to the conical dispersion, it is
not actually described by a Dirac Hamiltonian, as evidenced by the
fact that the cone is attached to an inseparable flat band
\cite{Green2010,conical_review}.  The underlying photonic crystals in
these studies are $\mathcal{T}$-symmetric, and so are the effective
media, which have NZ refractive index but no magneto-optical activity.

As we have noted, an unpaired Dirac point only exists if $\mathcal{T}$
is broken.  If the cone is centered at $\Gamma$, it may be described
by a NZ index effective medium; but unlike previously studied cases
\cite{CT2011, CT2012, dong2015, Ashraf2016, Jahani2016,xiao2017}, the
medium exhibits magneto-optical activity (non-vanishing imaginary
off-diagonal terms in the permeability tensor).  Intriguingly, the
enhancement of magneto-optical effects has previously been identified
as a promising application of NZ index media.  Engheta and co-workers
\citep{Davoyan2013, Davoyan2013oe, abbasi2015, EnghetaReview2017} have
noted that the strength of magneto-optical effects is characterized by
a Voigt parameter---the ratio of (imaginary) off-diagonal to (real)
diagonal components of the permeability or permittivity tensor---which
becomes large in NZ index media as the denominator becomes small.  We
indeed find strong magneto-optical activity in our effective medium,
with a near-unity effective Voigt parameter.  Another striking feature
of magneto-optical NZ index media is that they can support surface
magneto-plasmon-like \cite{Wang2015} waves that move unidirectionally
along a sample edge without being back-scattered by edge deformations
and other imperfections \citep{Davoyan2013, Davoyan2013oe,
  abbasi2015}.  The unidirectional and robust nature of these modes is
strongly reminiscent of the photonic topological edge states
\cite{lu2014} of 2D photonic crystals with topologically nontrivial
bandstructures, whose existence is guaranteed by topological
principles.  To our knowledge, no definite connection between these
two phenomena has previously been identified.  The present system
sheds light on the issue: the unpaired Dirac point coincides with a
topological transition, and as the topological band gap closes, the
topological edge states evolve continuously into the edge states of
the magneto-optical NZ index effective medium.  In a continuum
effective medium theory, such edge states can be characterized by the
Chern number as long as the Berry curvature is strongly localized to
the $\Gamma$ point, for example by the material dispersion or by
optimization of the full photonic band
structure~\cite{xiao2017,silveirinha2015,silveirinha2016}.

The unpaired Dirac point at $\Gamma$, like the non-Dirac band-crossing
points of Refs.~\cite{CT2011, CT2012, dong2015, Ashraf2016,
  Jahani2016}, is ``accidental'' in the sense that it must be realized
by fine-tuning the lattice; this is consistent with the fact that it
appears at a topological transition at finite $\mathcal{T}$-breaking.
By contrast, paired Dirac points in $\mathcal{T}$-symmetric lattices,
like graphene, can exist without fine-tuning if they are protected by
lattice symmetry.  (As explained below, however, one of our tuning
parameters is the applied magnetic field, which can be tuned
continuously in real experiments.)  Recently, a photonic system
containing an unpaired Dirac point has been proposed \cite{daniel1}
and observed \cite{Noh2017}, based on an array of helical optical
waveguides; however, this Dirac point exists in a Floquet
bandstructure defined via the paraxial equations of waveguide mode
evolution, which has no straightforward effective medium
interpretation.  In Ref.~\cite{ni2017}, it was shown that a
$\mathcal{T}$-broken photonic crystal can host an unpaired Dirac point
possessing many interesting behaviors, such as one-way Klein
tunneling.  This Dirac point occurs at a corner of the Brillouin zone
(similar to the Haldane model \cite{haldane1988}), rather than the
$\Gamma$ point.  In principle, such a Dirac point can be moved to
$\Gamma$ by additional fine-tuning.

\begin{figure}
  \centering
  \includegraphics[width=0.475\textwidth]{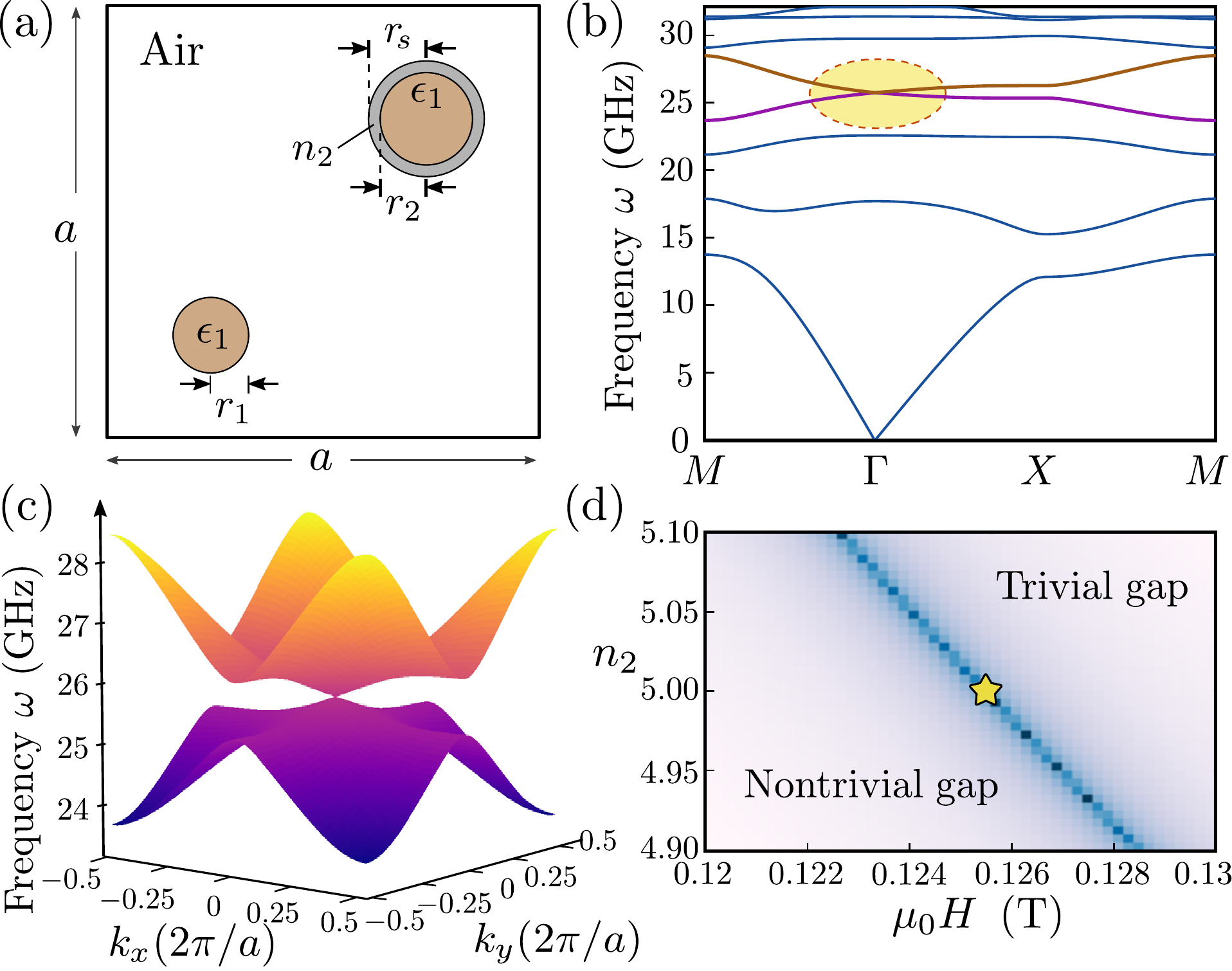}
  \caption{(a) Schematic of a 2D photonic crystal unit cell: a square
    of side $a$, containing two ferrite rods with radii $r_1$ and
    $r_2$, which form square sublattices.  One rod has a dielectric
    cladding with refractive index $n_2$ and outer radius $r_s$.  We
    set $a=4\,\textrm{cm}$, $r_1=0.088a$, $r_2=1.21\,r_1$, and
    $r_s=1.46\,r_1$.  The ferrite has $\epsilon_1=15$ and $\mu_0 M_s =
    0.191 \,\mathrm{T}$.  (b) TM photonic bandstructure for $n_2 = 5$
    and $\mu_0H = 0.125 \,\mathrm{T}$.  Bands 4 and 5 meet at an
    unpaired Dirac point at $\Gamma$.  (c) Bandstructure near the
    Dirac frequency ($\omega_D = 25.7\,\mathrm{GHz}$). (d) Heat map
    (in arbitrary units) showing the minimum frequency separation
    between bands 4 and 5, versus $n_2$ and $\mu_0H$.  The Dirac point
    occurs at a topological band transition; the Chern numbers for the
    4th band and 5th band are $-1$ and $1$ respectively in the
    nontrivial phase, and both 0 in the trivial phase.  The star
    indicates the parameters used in subplots (b) and (c).  }
  \label{unit_cell}
\end{figure}

We will study a photonic crystal based on gyromagnetic rods in a 2D
lattice.  Similar photonic crystals have previously been used to
realize photonic topological insulators \cite{raghu2008a,raghu2008b,
  wang2008, wang2009, Fu2010, Poo2011}; in those cases, the
$\mathcal{T}$-symmetric bandstructures contain band-crossing points
(either paired Dirac points, or quadratic band-crossing points
\cite{Chong2008}), and topologically nontrivial gaps are opened as
soon as $\mathcal{T}$ is broken by a biasing magnetic field $H \ne 0$
pointing in the out-of-plane ($\hat{z}$) direction.  By contrast, we
seek a topological transition occurring at a \textit{nonzero} value of
$H$.  To achieve this, we search a two-dimensional space spanned by
$H$ and a sublattice symmetry breaking parameter.  As shown in
Fig.~\ref{unit_cell}(a), we take the unit cell to be a square of
lattice constant $a$, with two ferrite rods lying along the diagonal,
surrounded by air.  The rods are separated by $a/\sqrt{2}$, and form
two square sublattices.  The have radii $r_1$ and $r_2$ respectively;
to further break the sublattice symmetry, we enclose the second rod in
a dielectric sheath with refractive index $n_2$ and outer radius
$r_s$.  The ferrite material has dielectric function $\epsilon_1$;
when biased by an out-of-plane magnetic field $H$, its magnetic
permeability tensor is \cite{Pozar}
\begin{equation}
  \overset\leftrightarrow{\mu} = \begin{bmatrix} \mu & i\alpha & 0 \\
    -i\alpha & \mu & 0 \\ 0 & 0 & \mu_0 \end{bmatrix},
  \label{permeability}
\end{equation}
where $\mu = 1 + \omega_m\omega_0/(\omega_0^2-\omega^2)$,
$\alpha = \omega_m\omega/(\omega_0^2 - \omega^2)$, $\omega_m = g \mu_0
M_s$, $\omega_0 = g \mu_0 H$, $g = 1.76\times10^{11}
\,\mathrm{Ckg}^{-1}$ is the gyromagnetic ratio, $M_s$ is the
saturation magnetization, and $\mu_0$ is the permeability of free
space.  Our specific parameter choices are listed in the caption of
Fig.~\ref{unit_cell}.

By tuning $H$ and other system parameters, we can generate a Dirac
point between bands 4 and 5, as shown in Fig.~\ref{unit_cell}(b)--(d).
The Dirac point appears at $\Gamma$ (i.e., $k_x = k_y = 0$), with
angular frequency $\omega_D \approx 25.7\,\mathrm{GHz}$, and is
unpaired.  (These and subsequent numerical results are obtained using
Comsol Multiphysics.)  Fig.~\ref{unit_cell}(d) plots part of the phase
diagram, showing that the Dirac point occurs along a topological phase
boundary where $H \ne 0$.  The gap is topologically nontrivial; the
calculated Chern numbers for the two bands are $\pm1$ below the
critical $H$, and 0 above.  As $H$ decreases further below the
depicted range, these two bands undergo another topological transition
(featuring a Dirac point at $X$) and re-open a trivial gap, consistent
with the principle that all gaps are topologically trivial for $H =
0$.

\begin{figure}
  \centering
  \includegraphics[width=0.47\textwidth]{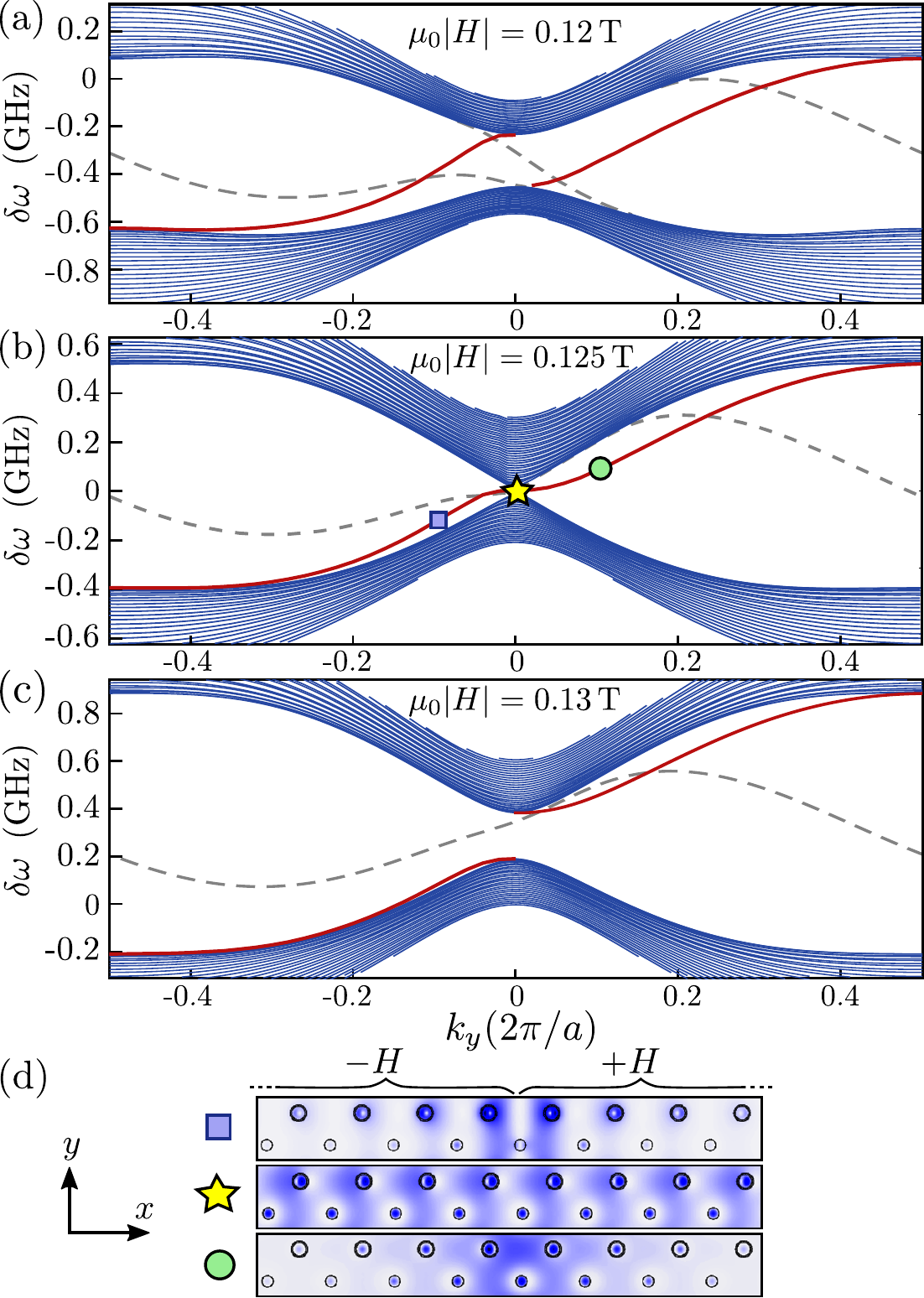}
  \caption{(a)--(c) Bandstructures for semi-infinite strip with two
    domains of opposite magnetic biases $\pm H$.  Results are shown
    for (a) $\mu_0|H| = 0.12\,\mathrm{T}$ (nontrivial gap), (b)
    $\mu_0|H| = 0.125\,\mathrm{T}$ (unpaired Dirac point), and (c)
    $\mu_0|H| = 0.13\,\mathrm{T}$ (trivial gap).  The angular
    frequencies are relative to $\omega_D = 25.7\,\mathrm{GHz}$.  The
    red curves indicate edge states localized at the interface between
    the domains; grey dashes are spurious edge states localized at the
    edges of the computational cell.  (d) Spatial distributions of
    field intensity $|E_z|^2$ for the edge states indicated by
    corresponding symbols in (b).  Only the 8 unit cells nearest to
    the domain wall are shown; the actual computational cell is 100
    unit cells wide along $x$.}
    \label{projected_dirac}
\end{figure}

The unpaired Dirac point is associated with topological edge states.
Fig.~\ref{projected_dirac}(a)--(c) shows bandstructures computed for a
strip that is periodic along $y$, and divided along $x$ into two
domains with opposite magnetic bias.  Below a critical field strength
$|H_c|$, there are two branches of gap-spanning edge states on the
domain wall, consistent with the Chern number difference of 2
[Fig.~\ref{projected_dirac}(a)].  At the transition, the edge states
merge continuously into a single branch which passes through the Dirac
point [Fig.~\ref{projected_dirac}(b)].  Above $|H_c|$, the edge states
no longer span the band gap [Fig.~\ref{projected_dirac}(c)].  At each
frequency, the edge states co-exist with bulk Dirac modes, but are
always centered at the domain wall, as shown in
Fig.~\ref{projected_dirac}(d).  Right at the Dirac point, the edge
state delocalizes while the bulk density of states goes to zero.

We now attempt to describe the Dirac cone using an effective medium
with permittivity $\tilde{\epsilon}$, and permeability tensor of the
form \eqref{permeability} with parameters $\tilde{\mu}$ and
$\tilde{\alpha}$.  For TM modes ($E_x = E_y = H_z = 0$), we can show
from Maxwell's equations that the bulk dispersion relation is
Dirac-like if the medium's frequency dependence satisfies
\begin{equation}
  \tilde{\epsilon} \tilde{\mu} \left(1-\frac{\tilde{\alpha}^2}{\tilde{\mu}^2}\right)
  = \left[\frac{1}{v_D}\; \frac{\delta\omega}{\omega}\right]^2,
  \label{dispersion}
\end{equation}
where $v_D$ is the Dirac speed, $\delta\omega \equiv \omega -
\omega_D$, and $\omega_D$ is the Dirac frequency \cite{supplemental}.

Now consider two uniform domains with equal and opposite Voigt
parameters $\tilde{\alpha}/\tilde{\mu}$, separated by a straight
domain wall parallel to $y$.  Maxwell's equations support TM modes
confined to the domain wall by the mismatch in
$\tilde{\alpha}/\tilde{\mu}$.  If Eq.~\eqref{dispersion} is obeyed,
these satisfy
\begin{align}
  |\delta \omega| &= v_D \sqrt{1-(\tilde{\alpha}/\tilde{\mu})^2} \, |k_y|
  \label{edge_dispersion} \\
  \gamma &= (\tilde{\alpha}/\tilde{\mu}) \, k_y, \label{gamma}
\end{align}
where $k_y$ is the wavenumber along the domain wall and $1/\gamma$ is
the penetration depth.  In order for the modes to decay away from the
domain wall, we require $\gamma > 0$.

\begin{figure}
  \includegraphics[width=0.49\textwidth]{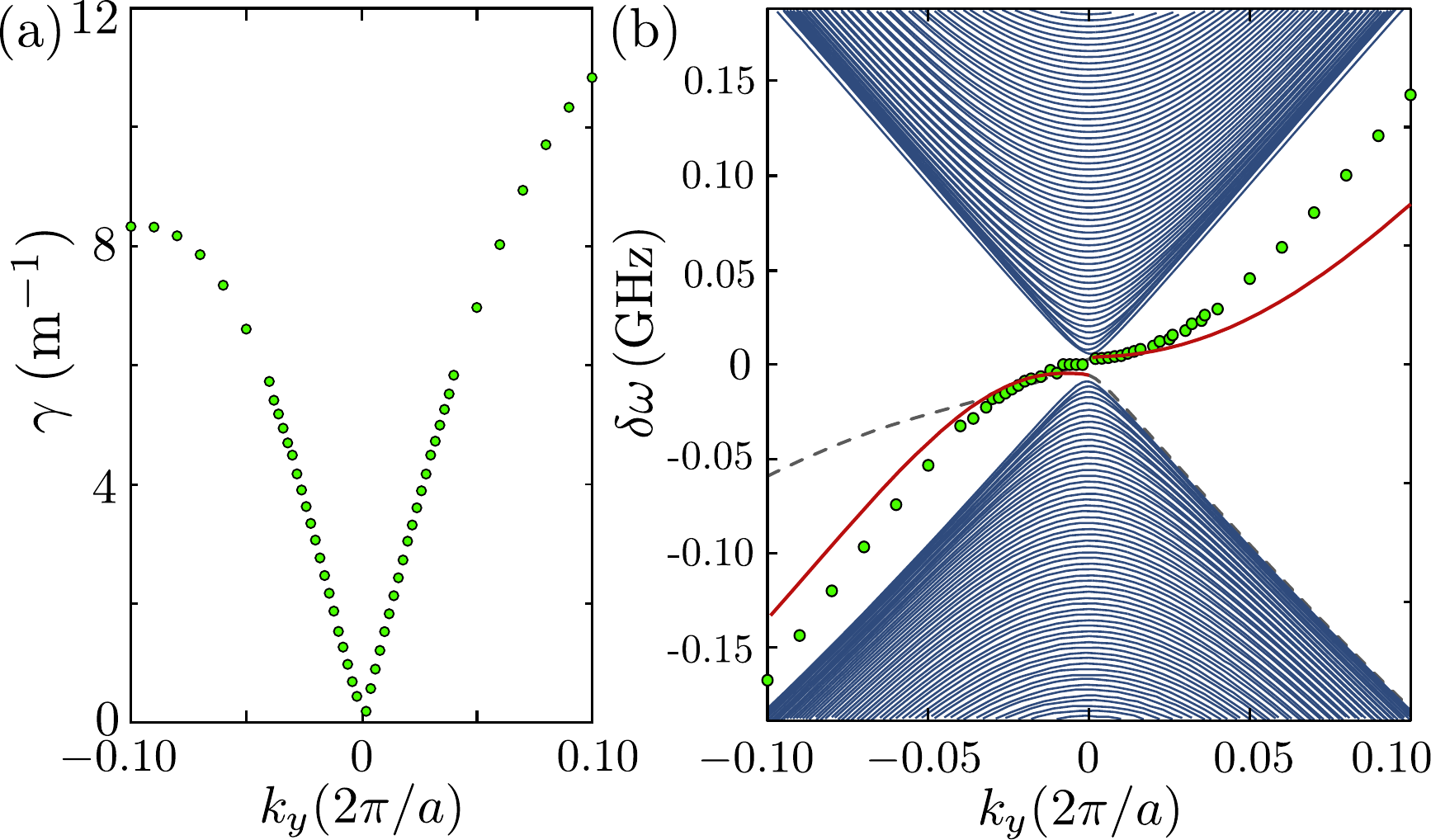}
  \caption{(a) Edge state decay parameter $\gamma$ versus $k_y$.
    These data are extracted from two-domain photonic crystal
    simulations by fitting $\langle|E_z|^2\rangle$ (averaged over each
    unit cell) to an $\exp(-2\gamma |x|)$ envelope.  (b) Close-up of
    the bandstructure near the Dirac point.  The solid curves show the
    numerically-calculated dispersion for the edge states (red) and
    bulk states (blue); the dashes are spurious edge states localized
    at edges of the computational cell, which is 300 unit cells wide
    with all other parameters the same as in
    Fig.~\ref{projected_dirac}(b).  The green dots show the effective
    medium prediction, calculated using the fitted value of $v_D$ and
    the decay parameters from (a). }
  \label{fitting}
\end{figure}

It should be noted that similar edge states can exist if
$\tilde{\epsilon}$, $\tilde{\mu}$, and $\tilde{\alpha}$ are
frequency-independent (so that Eq.~\eqref{dispersion} does
\textit{not} hold).  In that case, the edge dispersion relation
\eqref{edge_dispersion} is replaced by $\omega = v_0
\sqrt{1-(\tilde{\alpha}/\tilde{\mu})^2}\, |k_y|$, where $v_0$ is the
speed of light in the bulk.  Then $k_y\rightarrow 0$ at zero
frequency, rather than at a finite Dirac frequency $\omega_D$.

Eqs.~\eqref{dispersion}--\eqref{gamma} can be systematically fitted to
photonic crystal simulations.  From the bulk band diagram, we estimate
$v_D \approx 1.26\times10^{7} \,\textrm{ms}^{-1}$; the Dirac cone is
nearly isotropic, with fits along different crystal axes giving less
than $0.5\,\%$ variation in $v_D$.  Next, we extract $\gamma$ by
calculating the intensity profile $\langle|E_z|^2\rangle$ for the
photonic crystal edge states, averaging over each unit cell, and
fitting to an envelope $I_0 \exp(-2\gamma|x|)$.  The results are shown
in Fig.~\ref{fitting}(a).  Comparing this to Eq.~\eqref{gamma}, we
deduce that the Voigt parameter $\tilde{\alpha}/\tilde{\mu}$ switches
sign across $\omega_D$; very near $\omega_D$, it has almost constant
magnitude $|\tilde{\alpha}/\tilde{\mu}| \approx 0.94$.  From
Eq.~\eqref{dispersion}, this also implies that
$\tilde{\epsilon}\tilde{\mu} \rightarrow 0$ as
$\omega\rightarrow\omega_D$.  Finally, we use the fitted values of
$v_D$ and $\gamma$ together with
Eqs.~\eqref{edge_dispersion}--\eqref{gamma} to plot the edge state
dispersion curve $\delta \omega = \pm v_D \sqrt{k_y^2 - \gamma^2}$.
As shown in Fig.~\ref{edge_dispersion}(b), near the Dirac frequency
the dispersion curve predicted by the effective medium theory agrees
with the edge state dispersion curve from the photonic crystal
simulations.

The effective Voigt parameter of $|\tilde{\alpha}/\tilde{\mu}| \approx
0.94$ is close to the maximum allowed value; it cannot exceed unity,
as such a medium would lie in the ``Hall opacity'' regime
\cite{Davoyan2013} where no propagating bulk modes exist.  The actual
Voigt parameter within the underlying ferrite rods is $\alpha/\mu
\approx 1.53$ at the operating frequency.  However, the rods occupy
only $6\%$ of the photonic crystal's area.  An area-weighted
homogenization of the photonic crystal yields an effective Voigt
parameter of only $\approx 0.09$, while a different homogenization
scheme for magneto-optical media \cite{Abe1984} yields $\approx 0.2$
\cite{supplemental}.  Thus, the Voigt parameter of the Dirac
point-induced effective medium is substantially enhanced relative to
the homogenized photonic crystal, in agreement with previous arguments
that NZ index media can enhance magneto-optical activity
\cite{Davoyan2013, Davoyan2013oe, abbasi2015, EnghetaReview2017}.

\begin{figure}
  \centering
    \includegraphics[width=0.475\textwidth]{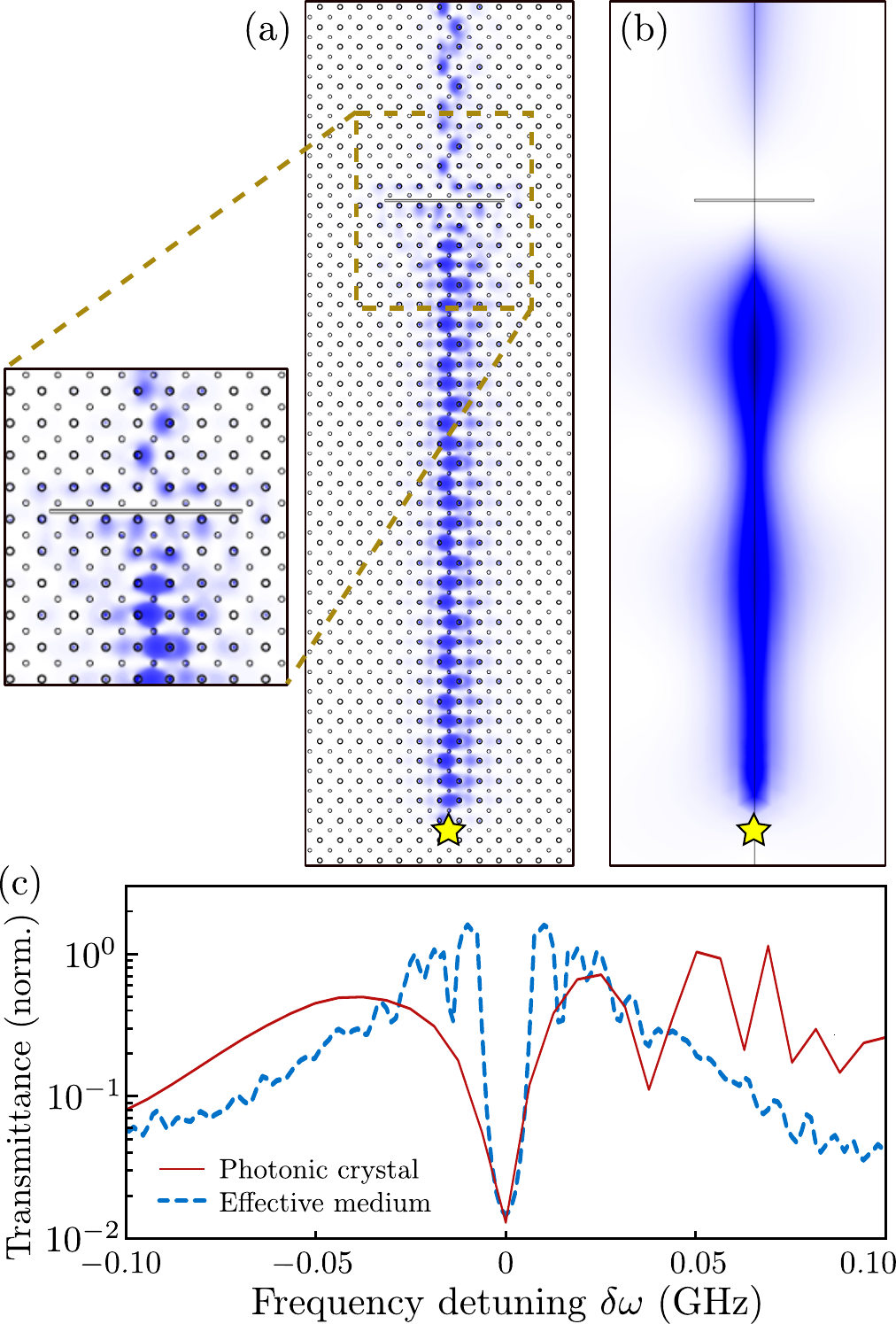}
      \caption{(a)--(b) Spatial distribution of the field intensity
        $|E_z|^2$, for surface waves moving along a domain wall with a
        perfect electrical conductor obstacle.  (a) Photonic crystal
        simulation, including an expanded view of the region near the
        obstacle.  (b) Effective medium simulation, using the
        effective medium fitting parameters from
        Fig.~\ref{fitting}(b).  The source positions are indicated by
        yellow stars.  In both cases, the detuning is $\delta \omega
        \approx 0.04\,\textrm{GHz}$.  For the effective medium,
        however, the Dirac frequency is taken to be $\omega_D \approx
        25.23\,\mathrm{GHz}$, a reduction of 1.8\% relative to the
        bulk bandstructure calculation; this value is determined from
        the transmittance dip in (c), and appears to be a
        discretization error.  (c) Transmittance versus frequency
        detuning $\delta \omega$, for the photonic crystal and the
        effective medium model.  The transmittance is estimated as the
        ratio of intensity $25a$ past the obstacle to the intensity
        $15a$ before the obstacle, along the domain wall.}
    \label{fig:comparison}
\end{figure}

Although the edge states delocalize ($\gamma \rightarrow 0$) at the
Dirac frequency, transmission along the edge appears to be quite
robust \textit{near} the Dirac frequency.  This is due to (i) the
reduction in the density of bulk states available to scatter into, and
(ii) the fact that the edge states are unidirectional and thus unable
to back-scatter.  Fig.~\ref{fig:comparison} shows the transmission
along a domain wall with a perfect electrical conductor obstacle.  The
simulations for both the photonic crystal and the effective medium
show a transmittance dip at the Dirac frequency, as shown in
Fig.~\ref{fig:comparison}(c).  At small detunings, the surface waves
are able to partially bypass the obstacle, as shown by the field
intensity plots in Fig.~\ref{fig:comparison}(a) for the photonic
crystal, and Fig.~\ref{fig:comparison}(b) for the effective medium.

In conclusion, we have designed a realistic gyromagnetic photonic
crystal with an unpaired Dirac point at $\Gamma$.  The Dirac medium
serves as a magneto-optical NZ index effective medium \cite{CT2011,
  CT2012, dong2015, Ashraf2016, Jahani2016,Davoyan2013, Davoyan2013oe,
  abbasi2015, EnghetaReview2017}, and we have identified the
``remnant'' of the topological edge state with the
magneto-plasmon-like surface states of the effective medium.  The
effective Voigt parameter is $\approx 0.94$, which is very near the
``Hall opacity'' regime \cite{Davoyan2013}, and enhanced by
approximately five-fold relative to the homogenized value.  We have
focused on the behavior of domain walls in the bulk of an otherwise
homogeneous photonic crystal. For finite arbitrarily-shaped domains,
however, the effective medium description of a photonic crystal is
known to be quite approximate, as it assumes the Bloch modes of
interest have a dominant Fourier component, which is typically not the
case for higher bands~\cite{smigaj2008}.  Moreover, if domain
boundaries do not preserve the bulk symmetries of the underlying
crystal, the effective medium must be complemented by additional
mode-matching boundary layers~\cite{simovski2011}.  In the future, it
would be interesting to identify models exhibiting unpaired Dirac
cones in lower bands, which might mitigate these limitations
\cite{CT2011}.

We are grateful to C.~T.~Chan for his stimulating and insightful
comments.  This work was supported by the Singapore MOE Academic
Research Fund Tier 2 Grant MOE2015-T2-2-008, the Singapore MOE
Academic Research Fund Tier 3 Grant MOE2016-T3-1-006, and the
Institute for Basic Science in Korea (IBS-R024-Y1).

\pagebreak
\begin{widetext}

\begin{center}
{\Large{Supplemental Material for:\\Realization of magneto-optical near-zero index medium by an
  unpaired Dirac point}}
\end{center}

\section{Derivation of effective medium parameters}

We seek an effective medium which can be mapped to the gyromagnetic
photonic crystal. Maxwell's equation in a gyromagnetic medium, with no
free charges or currents, are
\begin{align}
\nabla \times \textbf{E} &= i \omega \overset\leftrightarrow{\mu} \textbf{H} \label{Max1} \\
\nabla \times \textbf{H} &= -i \omega \epsilon \textbf{E} \label{Max2} \\
\nabla \cdot (\epsilon \textbf{E}) &= 0 \label{Max3} \\
\nabla \cdot (\mu \textbf{H}) &= 0. \label{Max4}
\end{align}
Here, $\epsilon$ is the scalar permittivity, and the permeability has
the form
\begin{equation}
  \overset\leftrightarrow{\mu} =
  \begin{bmatrix}
    \mu & i\alpha & 0 \\ -i\alpha & \mu & 0 \\ 0 & 0 & \mu_0
  \end{bmatrix}.
  \label{mutensor}
\end{equation}
Assume a uniform medium, so that both $\epsilon$ and
$\overset\leftrightarrow{\mu}$ are position-independent.  We consider
TM modes with $E_x = E_y = H_z = 0$.  For this polarization,
Eq.~(\ref{Max3}) is automatically satisfied.  Combining
Eq.~\eqref{Max1} with \eqref{mutensor} gives
\begin{align}
  - \nabla^2_{\mathrm{2D}} E_z &=
  i \omega  \left[\mu(\partial_x H_y - \partial_y H_x)
    - i\alpha(\partial_x H_x + \partial_y H_y)\right],
  \label{intermed1}
\end{align}
where $\nabla^2_{\mathrm{2D}} \equiv \partial_x^2 + \partial_y^2$ is
the 2D Laplacian.  Next, combining Eq.~\eqref{Max4} with
\eqref{mutensor} gives
\begin{equation}
  \mu\left(\partial_x H_x + \partial_y H_y\right)
  = -i\alpha \left(\partial_x H_y - \partial_y H_x\right).
  \label{intermed2}
\end{equation}
Combining Eqs.~\eqref{intermed1}--\eqref{intermed2} and
Eq.~\eqref{Max2} yields
\begin{equation}
  \left[\nabla^2_{\mathrm{2D}} + \epsilon 
    \mu\left(1-\frac{\alpha^2}{\mu^2}\right)
    \omega^2\right] \, E_z = 0.
  \label{intermed3}
\end{equation}
So far, we have not made any assumptions about the frequency
dispersion of the medium.  We now assume that the bulk dispersion
obeys the Dirac dispersion relation
\begin{equation}
  \omega = \omega_D + v_D |\mathbf{k}|,
  \label{diracdisp}
\end{equation}
where $\omega_D$ is the Dirac frequency, $v_D$ is the effective speed
of the Dirac modes, and $\mathbf{k} = (k_x , k_y)$ is the 2D
in-plane wavevector.  Comparing
Eqs.~\eqref{intermed3}--\eqref{diracdisp}, we deduce that
\begin{equation}
  \epsilon \mu \left(1-\frac{\alpha^2}{\mu^2}\right)
  = \left[\frac{1}{v_D}\; \frac{\delta\omega}{\omega_D}\right]^2,
  \label{diracfit}
\end{equation}
where $\delta \omega \equiv \omega - \omega_D$ (we assume throughout
that $|\delta\omega| \ll \omega_D$).  This means that the effective
refractive index vanishes for $\delta \omega \rightarrow 0$; this is a
NZ index medium.

Next, consider two adjacent domains, separated by a domain wall at $x
= 0$.  The magnetic bias has opposite signs for $x > 0$ and $x < 0$.
In other words, for $x < 0$ we replace $\alpha$ with $-\alpha$ in
Eq.~\eqref{mutensor}.  We now seek solutions of the form
\begin{equation}
  E_{z} = E_0 \, e^{-\gamma |x| + ik_y y}.
\end{equation}
These modes are localized around $x = 0$, with wavenumber $k_y$
parallel to the domain wall and penetration depth $1/\gamma$.  Note
that $E_z$ is necessarily continuous across the domain wall.
Substituting into Eq.~\eqref{intermed3} yields the condition
\begin{equation}
  k_y^2 - \gamma^2 = \epsilon\mu \left(1-\frac{\alpha^2}{\mu^2}\right)\, \omega^2.
  \label{intermed4}
\end{equation}
Moreover, knowing $E_z$ we can use Eq.~\eqref{Max1} to retrieve the
magnetic fields:
\begin{align}
  \mathbf{H} &= \frac{1}{i\omega}\; \overset\leftrightarrow{\mu}^{-1} \nabla \times\mathbf{E} \\
  \begin{bmatrix}H_x \\ H_y \end{bmatrix}
  &= \frac{E_0}{i\omega} \; \overset\leftrightarrow{\mu}^{-1} \begin{bmatrix}ik_y \\ \pm \gamma
  \end{bmatrix} \, e^{\mp\gamma x + ik_y y} \\
  &= \frac{E_0}{i\omega(\mu^2-\alpha^2)} \begin{bmatrix}i(\mu k_y-\alpha\gamma) \\ \pm (\mu\gamma-\alpha k_y)
  \end{bmatrix} e^{\mp\gamma x + ik_y y},
\end{align}
where the $\pm$ signs denote the $x > 0$ and $x < 0$ domains
respectively.  Due to the absence of surface currents, $H_y$ must also
be continuous across the domain wall, and hence
\begin{equation}
  \gamma = \frac{\alpha k_y}{\mu}.
  \label{penetration}
\end{equation}

The parameter $\gamma$ must be positive.  Eq.~\eqref{penetration}
implies that if a domain wall state exists for $k_y > 0$, then
$\alpha/\mu > 0$ for $k_y > 0$.  Conversely, if a domain wall state
exists for $k_y < 0$, then $\alpha/\mu < 0$ for $k_y <0$.  In
Fig.~2(b) and Fig.~3 of the main text, we observe that domain wall
states of the photonic crystal exist for both positive and negative
$k_y$.  Moreover, the domain wall state's relative frequency
$\delta\omega \equiv \omega - \omega_D$ switches sign as $k_y$
switches sign.  This implies that $\alpha/\mu$ switches sign across
the Dirac frequency.

Using Eq.~\eqref{penetration}, we can simplify Eq.~\eqref{intermed4}
to $\epsilon\mu\omega^2 = k_y^2$.  Applying the Dirac medium condition
\eqref{diracfit} then yields the dispersion relation for the domain
wall states:
\begin{equation}
  \delta\omega^2 = v_D^2 \left[1-\left(\frac{\alpha}{\mu}\right)^2\right]
  k_y^2.
  \label{edgedisp}
\end{equation}

We now estimate $v_D$ by taking the dispersion along two directions,
$\Gamma$-$X$ and $\Gamma$-$M$.  The values of $v_D$ and $\omega_D$
estimated for both cases are almost identical, and we take the mean
values $v_D = 1.255\times10^{7} \,\textrm{ms}^{-1}$ and $\omega_D =
25.70\,\mathrm{GHz}$.  By fitting the penetration constant of the
domain wall states to Eq.~\eqref{penetration}, we find $|\alpha/\mu|
\approx 0.961$ near $\omega_D$ [see Fig.~3(a) of the main text].  This
implies, via Eq.~\eqref{edgedisp}, that near $\omega_D$ the domain
wall states have an approximately linear dispersion relation, with
group velocity
\begin{equation}
  v_s = v_D \sqrt{1-\frac{\alpha^2}{\mu^2}}.
\end{equation}
Since $|\alpha/\mu|$ is close to unity, $v_s \ll v_D$.

\section{Effective Voigt parameter under crude homogenization}

In the main text, we point out that the effective medium's effective
Voigt parameter of $|\tilde{\alpha}/\tilde{\mu}| \approx 0.96$ is
remarkably large, given that (i) the actual Voigt parameters in the
ferrite rods is $\alpha/\mu \approx 1.5$ at the operating frequency,
and (ii) the rods occupy only $6\%$ of the photonic crystal's area
(see Fig.~1 of the main text).  Crudely weighting $\alpha/\mu$ by area
yields an effective Voigt parameter of approximately $0.092$.

A slightly more sophisticated scheme for homogenizing magneto-optic
media was derived in Ref.~\cite{Abe1984}.  That paper considered
cylinders of gyroelectric material, with dielectric tensors of the
form
\begin{eqnarray}
  \overset\leftrightarrow{\epsilon_2} =
  \begin{bmatrix}
    \epsilon_2 & i\gamma & 0 \\ -i\gamma & \epsilon_2 & 0 \\ 0 & 0 & \epsilon_2
  \end{bmatrix}.
\end{eqnarray}
The background medium has permittivity $\epsilon_1$, and the magnetic
permeability is unity throughout.  It was found that the homogenized
effective dielectric parameters are
\begin{align}
  \tilde{\epsilon}_x &= \tilde{\epsilon}_y =
  \epsilon_1 + f(\epsilon_2-\epsilon_1)\left[1 +\frac{(1-f)(\epsilon_2-\epsilon_1)}{2\epsilon_1}\right]^{-1}\\
  \tilde{\epsilon}_z &= (1-f)\epsilon_1 + f\epsilon_2\\
  \tilde{\gamma} &=  \gamma f \left[
    1 +\frac{(1-f)(\epsilon_2-\epsilon_1)}{2\epsilon_1}\right]^{-1},
\end{align}
where $\tilde{\epsilon}_x$, $\tilde{\epsilon}_y$, and
$\tilde{\epsilon}_z$ are the on-diagonal components, and
$\tilde{\gamma}$ is the magneto-optical off-diagonal component of the
effective permittivity tensor, and $f$ is the volume fraction of the
gyroelectric component.

We adapt this calculation to our gyromagnetic photonic crystal by the
mapping
\begin{equation}
  \left\{\begin{array}{rl}
  \nabla\times \mathbf{E} &= i \omega \mathbf{B} \\
  \nabla\times \mathbf{B} &= -i\omega \overset\leftrightarrow{\epsilon} \mathbf{E}\\
  \mu &= 1
  \end{array}
  \right\} \;\;\leftrightarrow\;\;
  \left\{\begin{array}{rl}
  \nabla\times \mathbf{B} &= -i\omega \mathbf{E}\\
  \nabla\times \mathbf{E} &= i \omega \overset\leftrightarrow{\mu} \mathbf{H} \\
  \epsilon &= 1
  \end{array}
  \right\}
\end{equation}
As this is only intended as a rough comparison, we ignore the
non-unity permittivities of the ferrite rods.  At the operating
frequency of $\omega \approx \omega_D$, the ferrite permeability is
\begin{eqnarray}
  \overset\leftrightarrow{\mu} =
  \begin{bmatrix}
    \mu & i\alpha & 0 \\ -i\alpha & \mu & 0 \\ 0 & 0 & \mu_0
  \end{bmatrix}, \;\;\;
  \mu \approx -3.2, \;\; \alpha \approx -4.9.
  \label{mutensor2} 
\end{eqnarray}
(See the parameters in the main text.)  The volume fraction is $f
\approx 0.06$.  Thus,
\begin{eqnarray}
\tilde{\mu} = 1.2, \;\; \tilde{\alpha} = 0.3.
\end{eqnarray}
This corresponds to a Voigt parameter of $\tilde{\alpha}/\tilde{\mu}
\approx 0.23$.
\end{widetext}

\end{document}